\documentclass[12pt]{article}
\pdfoutput=1
\usepackage{putex}
\usepackage{graphicx}
\usepackage{caption}
\usepackage{subcaption}
\usepackage{epstopdf}
\usepackage{enumerate}
\usepackage{cite}
\usepackage{youngtab}
\usepackage{braket}
\usepackage{bbm}
\usepackage{tensor}
\usepackage{slashed}
\usepackage[aligntableaux=center]{ytableau}
\usepackage{multirow}

\newcommand{\vphi}{\varphi}

\newcommand{\abs}[1]{\left\lvert #1 \right\rvert}

\newcommand {\be} {\begin {equation}}
\newcommand {\ee} {\end {equation}}

\newcommand {\bes} {\begin {equation*}}
\newcommand {\ees} {\end {equation*}}

\newcommand{\es}[2] {\begin{equation} \label{#1} \begin{split} #2 \end{split} \end{equation}}

\newcommand{\Z}{\mathbb{Z}}

\newcommand{\R}{\mathbb{R}}
\newcommand{\C}{\mathbb{C}}

\newcommand{\cO}{{\cal O}}

\newcommand{\tQ}{\widetilde{Q}}

\def\Tr{\mop{Tr}}

\newcommand{\beq}{\begin{equation}}
\newcommand{\eeq}{\end{equation}}

\begin{document}

\preprint{PUPT-2532}

\institution{PU}{Joseph Henry Laboratories, Princeton University, Princeton, NJ 08544, USA}

\title{A New Duality Between $\mathcal{N}=8$ Superconformal Field Theories in Three Dimensions}

\authors{Nathan B.~Agmon, Shai M.~Chester, and Silviu S.~Pufu}

\abstract{
We propose a new duality between two 3d $\mathcal{N}=8$ superconformal Chern-Simons-matter theories:  the $U(3)_1 \times U(3)_{-1}$ ABJM theory and a theory consisting of the product between the $\left(SU(2)_3\times SU(2)_{-3}\right)/\mathbb{Z}_2$ BLG theory and a free ${\cal N} = 8$ theory of eight real scalars and eight Majorana fermions.    As evidence supporting this duality, we show that the moduli spaces, superconformal indices, $S^3$ partition functions, and certain OPE coefficients of BPS operators in the two theories agree.  
}
\date{\today}

\maketitle

\tableofcontents

\section{Introduction and Summary}

\begin{figure}[tbp]
\begin{center}
   \def\svgwidth{.4\textwidth}
  {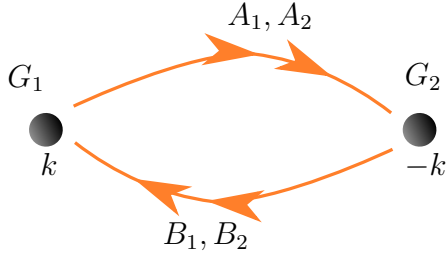}
\caption{The field content of the two-gauge group description of ${\cal N} = 8$ SCFTs.  The gauge group is $G_1 \times G_2$ with opposite Chern-Simons levels for the two factors.  The matter content consists of two pairs of bifundamental chiral multiplets whose bottom components are denoted by $A_1$, $A_2$ and $B_1$, $B_2$.  As explained in the main text, such theories have ${\cal N} = 8$ SUSY at the IR fixed point only for special values of $k$ and/or for special gauge groups $G_1$ and $G_2$.}
\label{TwoNodeFigure}
\end{center}
\end{figure}

Maximally supersymmetric (${\cal N} = 8$) superconformal field theories in three dimensions have received quite a bit of attention due to their interpretation as M2-brane theories and due to the many new exact supersymmetric localization results that have allowed for several precision tests of AdS/CFT\@.  While some of these theories have several distinct microscopic descriptions, they can all be described by a few infinite families of Chern-Simons (CS) theories with a product gauge group coupled to two pairs of matter chiral multiplets transforming in the bifundamental representation of the gauge group---see Figure~\ref{TwoNodeFigure}. These families are:  
 \begin{itemize}
  \item BLG theories:  These are $SU(2)_k \times SU(2)_{-k}$ (denoted BLG$_k'$) and $(SU(2)_k \times SU(2)_{-k})/\mathbb{Z}_2$ (denoted BLG$_k$) gauge theories, which preserve manifest ${\cal N} = 8$ supersymmetry for any integer Chern-Simons level $k$.    This description of the BLG theories is a reformulation \cite{VanRaamsdonk:2008ft, Bandres:2008vf} of the original work of Bagger, Lambert,  \cite{Bagger:2007vi,Bagger:2007jr,Bagger:2006sk} and Gustavsson \cite{Gustavsson:2007vu} (BLG).  While these theories were originally believed to have an interpretation as effective theories on $2$ coincident M2-branes, and this is indeed true for certain small values of $k$, their M-theory interpretation, if any, is still unknown for arbitrary $k$.
   \item  ABJM or ABJ theories:  These are $U(N)_k \times U(M)_{-k}$ gauge theories (denoted ABJM$_{N, k}$ if $N=M$ and ABJ$_{N, M, k}$ if $N \neq M$), which are believed to flow to IR fixed points with ${\cal N} = 8$ supersymmetry only if the Chern-Simons level is $k=1$ or $2$ and $\abs{N-M} \leq k$.  The theories with $M=N$ were first introduced by Aharony, Bergman, Jafferis, and Maldacena (ABJM) in \cite{Aharony:2008ug}, and those with $M \neq N$ by Aharony, Bergman, and Jafferis (ABJ) in \cite{Aharony:2008gk}.  These theories can be interpreted as effective theories on $N$ coincident M2-branes placed at a $\C^4/\Z_k$ singularity in the transverse directions.  Due to the dualities \cite{Aharony:2008ug,Bashkirov:2011pt}
   \es{DualityABJ}{
    \text{ABJ}_{N+1, N, 1} \quad &\cong \quad \text{ABJM}_{N, 1} \,, \\
    \text{ABJ}_{N+2, N, 2} \quad &\cong \quad \text{ABJM}_{N, 2}  \,,
   }
the only independent theories in this family are the ABJM$_{N, 1}$, ABJM$_{N, 2}$, and ABJ$_{N+1, N, 2}$ theories.
\end{itemize}

The case of the ABJM$_{1, 1}$ theory is worth noting:  this theory is equivalent to a free theory of $8$ massless real scalars and $8$ massless Majorana fermions.  This is the theory on one M2-brane in flat space, where the $8$ scalars parameterize the location of the brane in the transverse space.  The case ABJM$_{N, 1}$ for $N>1$, which corresponds to a stack of $N$ M2-branes in flat space, flows to a product of two decoupled CFTs in the infrared (see for instance \cite{Bashkirov:2010kz}).  One of these CFTs is free (and equivalent to the ABJM$_{1, 1}$ theory), and corresponds to the center-of-mass motion of the stack of branes.  The other CFT in the product is interacting and strongly coupled.

In addition to the dualities between ABJM / ABJ theories already mentioned, there are further dualities that relate the BLG and ABJM theories at certain small values of $k$.  For instance \cite{Lambert:2010ji,Bashkirov:2011pt}: 
 \es{oldDualsBLG}{
 \text{BLG$_1$} \quad &\cong \quad \text{ABJM$_{2, 1}$} \,, \\
 \text{BLG$_2'$} \quad &\cong \quad \text{ABJM$_{2, 2}$} \,, \\
  \text{BLG$_4$} \quad &\cong \quad \text{ABJ$_{2, 3, 2}$} \,.
 }
Furthermore, it is possible to conjecture other dualities that come from the fact that the $k=1, 2$ ABJM and the $k=2$ ABJ theories can be thought of as the IR limits of the maximally supersymmetric Yang-Mills theory with gauge algebra $\mathfrak{u}(N)$, $\mathfrak{so}(2N)$, and $\mathfrak{so}(2N+1)$, respectively  \cite{Kim:2009wb,Kapustin:2010xq,Gang:2011xp,Bianchi:2012ez}.   At small $N$, there are various coincidental isomorphisms between these Lie algebras, which themselves induce isomorphisms between the corresponding ${\cal N} = 8$ SCFTs.  For instance, since $\mathfrak{u}(2) \cong \mathfrak{u}(1) \oplus \mathfrak{so}(3)$, one expects that the ABJM$_{2, 1}$ theory should  be isomorphic to the product between the  ABJM$_{1, 1}$ theory and the ABJ$_{2, 1, 1}$ theory.

The purpose of this paper is to present yet another duality between the ABJ(M) and BLG theories that is not included in the list above.  It is:
 \es{NewDuality}{
  \boxed{ \text{BLG$_3$ $\otimes$ ABJM$_{1, 1}$} \quad \cong \quad \text{ABJM$_{3, 1}$ .}}
 }
Recalling that the ABJM$_{3, 1}$ theory has a decoupled free sector isomorphic to ABJM$_{1, 1}$ theory as well as an interacting one, this duality can be rephrased as  
 \es{NewDualityRephrase}{
  \boxed{ \text{BLG$_3$} \quad \cong \quad \text{interacting sector of ABJM$_{3, 1}$ .}}
 }
Thus, our new duality \eqref{NewDuality}--\eqref{NewDualityRephrase} provides an interpretation for the BLG theory at level $k=3$:  it is the interacting sector of the theory on three coincident M2-branes.  Quite curiously, this duality casts the $k=3$ BLG theory as a theory on three coincident M2-branes, unlike the original intuition that BLG theories should be related to theories on two M2-branes.

It is worth mentioning that the ${\cal N} = 8$ SCFTs mentioned above may have other descriptions that are not two-node gauge quivers.  An important example is that the ABJM$_{N, 1}$ theory has the same IR fixed point as an ${\cal N} = 4$ $U(N)$ gauge theory with a fundamental hypermultiplet and an adjoint hypermultiplet \cite{Bashkirov:2010kz,Bashkirov:2011pt}.  In fact, it is this latter description that we will use in some of our computations in the ABJM$_{3, 1}$ theory that we perform in order to check \eqref{NewDuality}--\eqref{NewDualityRephrase}.

\begin{figure}[t!]
\begin{center}
   \includegraphics[width=0.49\textwidth]{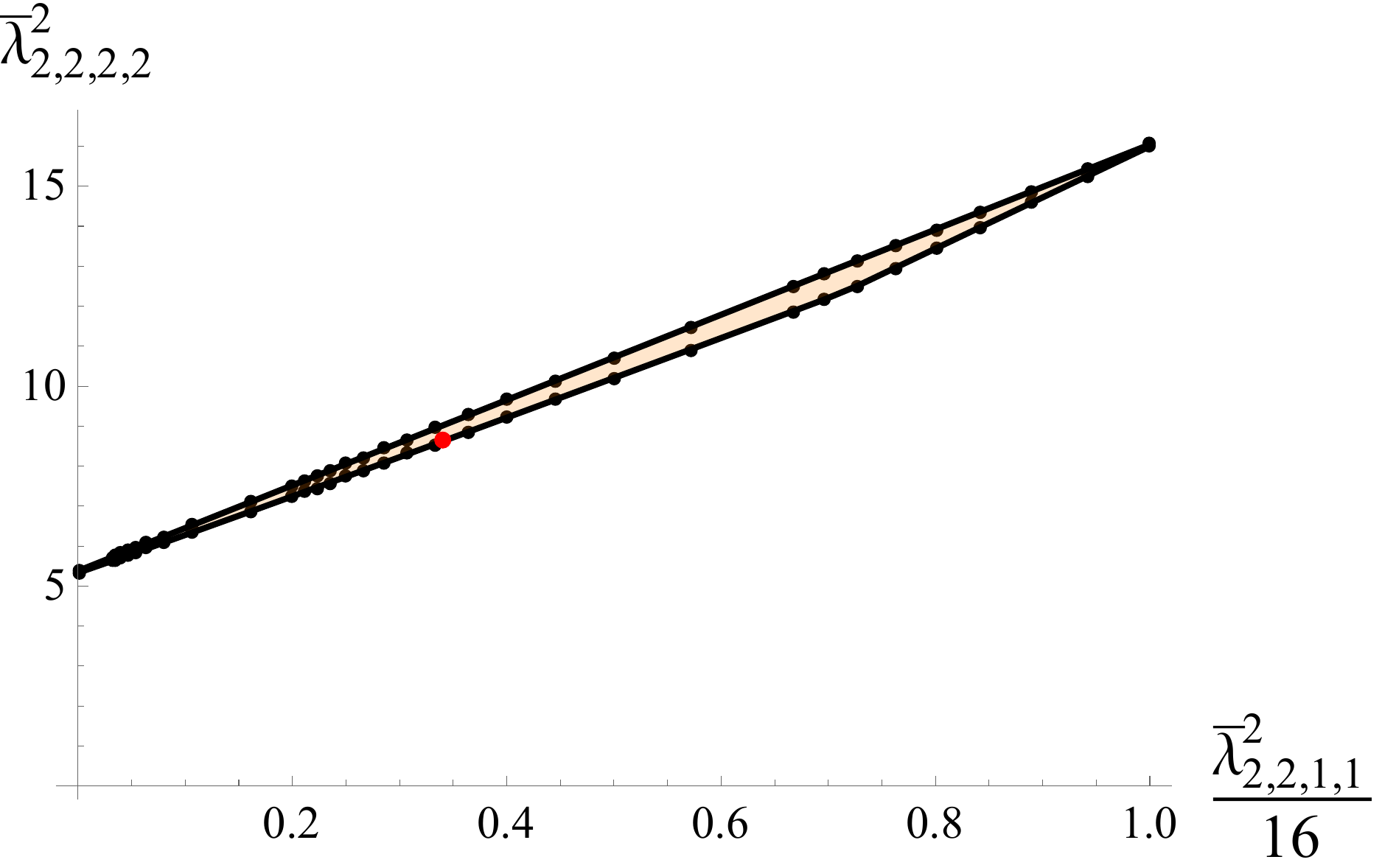}
   \includegraphics[width=0.5\textwidth]{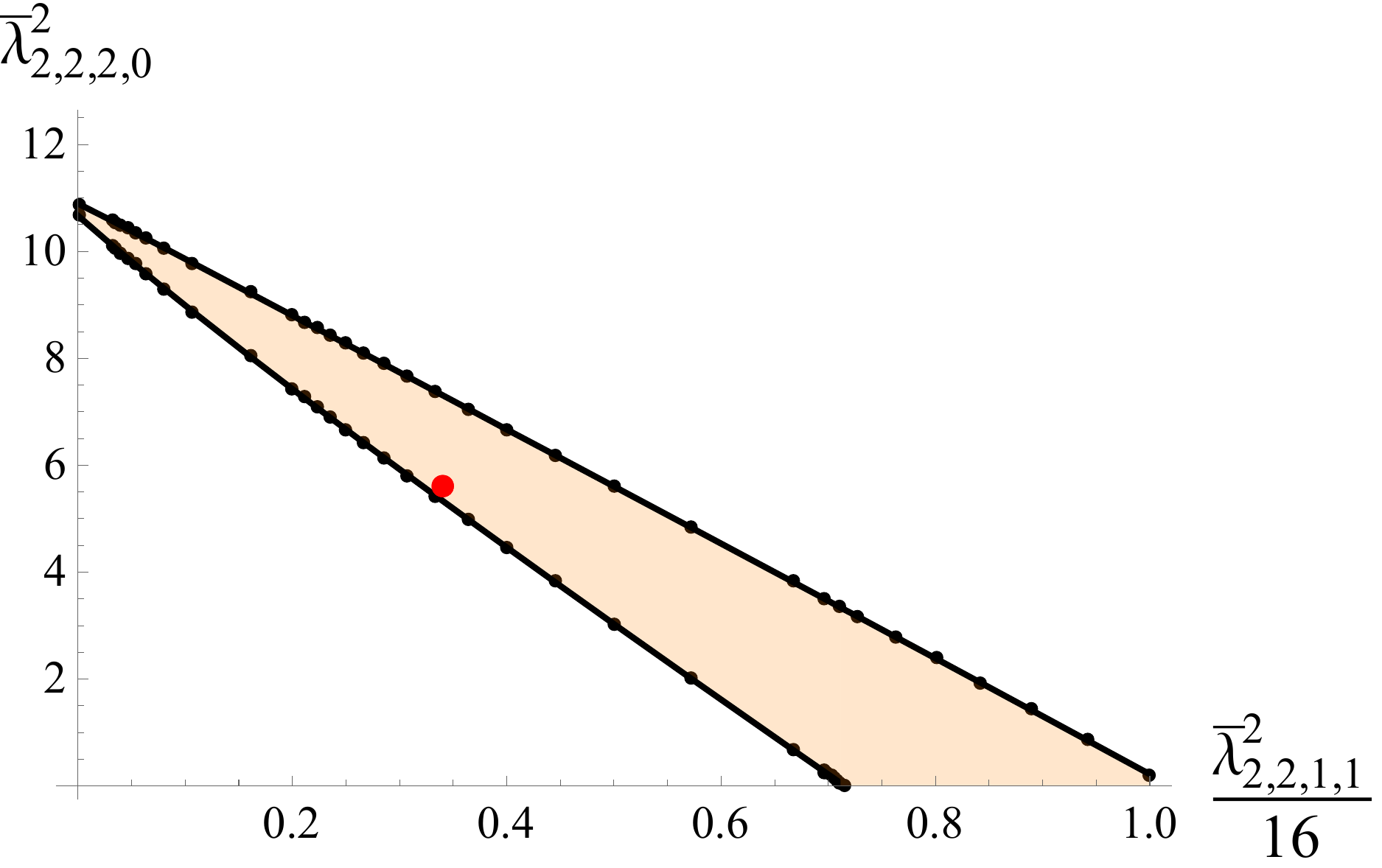}
\caption{Upper and lower bounds on $\bar\lambda_{2,2,2,2}^2$  and $\bar\lambda_{2,2,2,0}^2$ OPE coefficients in terms of the stress tensor OPE coefficient $\bar\lambda_{2,2,1,1}^2$, where the orange shaded regions are allowed, and the plot ranges from the supergravity limit $\bar\lambda_{2,2,1,1}\to0$ ($c_T\to\infty$) to the free theory $\bar\lambda_{2,2,1,1}^2=16$. The red dots denote the values for the interacting sector of the ABJM$_{3, 1}$ theory or for the BLG$_3$ theory, given in \eqref{OPEFinal}. The $\bar\lambda_{2,2,2,2}^2$ bounds can be mapped into the $\bar\lambda_{2,2,2,0}^2$ bounds using \eqref{crossConstraints}. These bounds were computed following \cite{Chester:2014fya}, except with the improved parameters $j_\text{max}=88$ and $\Lambda=43$.}
\label{BBounds}
\end{center}
\end{figure}

As was checked in previous dualities, for our proposed duality we match the moduli spaces and superconformal indices on each side of the duality. We also match the values of the $S^3$ partition functions of the two theories. In addition, we provide a new check using the recently proposed supersymmetric localization of 3d $\mathcal{N}=4$ theories to a topological 1d sector \cite{Dedushenko:2016jxl}. Using this method, we calculate the two- and three-point functions of low-lying half and quarter BPS operators, which we use to extract their OPE coefficients, listed in \eqref{OPEcoeff}, \eqref{OPEcoeffs2}, and \eqref{OPEcoefficentsAll}. For the OPE coefficients in the four point function of the stress tensor, we compare these values to the conformal bootstrap bounds of \cite{Chester:2014fya} in Figure \ref{BBounds}. We find that the exact values come close to saturating the lower bounds.

The paper is organized as follows. In Section~\ref{review} we review ABJM$_{3, 1}$ and BLG$_3$ theories and demonstrate the explicit operator matching for low-lying BPS operators, including matching the superconformal index. In Section~\ref{modSpace} we match the moduli spaces. In Section~\ref{S3partition} we compute and match the values of the $S^3$ partition functions. In Section~\ref{topStuff} we study certain 1d topological sectors of each theory, and extract the OPE coefficients of low-lying BPS operators. In Appendix~\ref{4point} we discuss the four point function, and in Appendix~\ref{More} we list more OPE coefficients. 

\section{Operator Spectrum and the Superconformal Index}
\label{review}

\subsection{Low-lying BPS Operator Spectrum} 

Let us start by introducing the two ${\cal N}  =8$ theories we argue are dual in more detail and compare their operator spectra.  These theories belong to the family of $\mathcal{N}=6$ Chern-Simons-matter theories  \cite{Aharony:2008ug} that have gauge group $U(N)\times U(N)$ or $(SU(N)\times SU(N))/\mathbb{Z}_N$ with Chern-Simons coefficients $k$ and $-k$ for the two gauge groups. In $\mathcal{N}=2$ notation, the matter content consists of chiral multiplets with scalar components $A_1,A_2$ and $B_1,B_2$ that transform under the product gauge group as $(\bold{N},\overline{\bold{N}})$ and $(\overline{\bold{N}},\bold{N})$, respectively. The theories have a quartic superpotential
\es{superpot}{
W=\frac{2\pi}{k}\epsilon^{ab}\epsilon^{\dot{a}\dot{b}}\Tr(A_aB_{\dot{a}}A_bB_{\dot{b}}) \,,
}
which preserves an $SU(2)\times SU(2)$ flavor symmetry under which $A_a$ transforms as $(\bold2,\bold1)$ and  $B_{\dot{a}}$ transforms as $(\bold1,\bold2)$. These theories also have a manifest $SU(2)_R$ symmetry (corresponding to ${\cal N} = 3$ SUSY) under which $(A_a, B_{\dot a}^\dagger)$ form doublets, a $U(1)_R$ subgroup of which being the ${\cal N} = 2$ R-symmetry under which $A_a$ and $B_{\dot a}$ have canonical R-charge $1/2$.  The $SU(2) \times SU(2)$ flavor symmetry combines with the $SU(2)_R$ symmetry to form an $SU(4)_R$ R-symmetry, as appropriate for ${\cal N} = 6$ SUSY\@.  

The theories with $U(N) \times U(N)$ gauge group have an additional topological $U(1)_T$ symmetry under which only monopole operators are charged.  When $k=1$ or $2$, one can find additional R-symmetry generators, which together with $SU(4)_R$ and $U(1)_T$ combine into an $SO(8)_R$ symmetry; these theories thus have ${\cal N} = 8$ SUSY\@.  We will of course be interested only in the case $N=3$, $k=1$.  The theories with $(SU(N)\times SU(N))/\mathbb{Z}_N$ gauge group in general do not have a similar R-symmetry enhancement.  When $N=2$, however, one can show that because $A_a$ and $B_{\dot a}$ now transform in the same gauge representation, the superpotential \eqref{superpot} has an $SU(4)$ flavor symmetry (which contains the $SU(2) \times SU(2)$ flavor symmetry as well as a baryonic symmetry $U(1)_t$ under which the $A_a$ have charge $+1$ and the $B_{\dot a}$ have charge $-1$), which combines with the $SU(2)_R$ symmetry mentioned above to form an $SO(8)_R$ R-symmetry.  Such an enhancement occurs for any $k$, but we will focus on the case $k=3$.

We would like to compare the operator spectra of the $U(3)_1 \times U(3)_{-1}$ (ABJM$_{3, 1}$) and $(SU(2)_3 \times SU(2)_{-3}) / \Z_2$ (BLG$_3$) theories.  We will do so by explicitly constructing various operators in short representations of the superconformal algebra, so let us review briefly what kinds of representations are possible.  (For more details, see~\cite{Dolan:2008vc}.)   In any ${\cal N} = 8$ SCFT, operators transform in unitary representations of the superconformal algebra $\mathfrak{osp}(8|4)$.  These representations are of two types, customarily denoted by $A$ and $B$, a number between $0$ and $3$ or a sign:  the $A$-type ones are $(A, 0)$, $(A, 1)$, $(A, 2)$, $(A, 3)$, $(A, +)$, $(A, -)$ and the $B$-type ones are $(B, 0)$, $(B, 1)$, $(B, 2)$, $(B, 3)$, $(B, +)$, $(B, -)$.  In addition, we also need to specify the $SO(8)_R$ representation of the superconformal primary, whose Dynkin labels we write as a superscript.  The representations of $B$-type are shorter than the corresponding $A$ representations.  The longest representations are of $(A, 0)$ type and do not obey any shortening condition;  the shortest representations ($1/2$-BPS) are of $(B, +)$ or $(B, -)$ type.  One of the latter, after choosing the $SO(8)_R$ triality convention, can be taken to be $(B, +)^{[0020]}$, which contains the stress-energy tensor and must be present in all local ${\cal N} = 8$ SCFTs\@.   Here $[0020]$ are the Dynkin labels of the ${\bf 35}_c$ irrep of $SO(8)_R$, in which the superconformal primary of the stress tensor representation transforms.   (Our choice of $SO(8)_R$ triality is uniquely specified by requiring that the eights supercharges transform in the ${\bf 8}_v$ and that the superconformal primary of the stress tensor multiplet transforms in the ${\bf 35}_c$.)

We will exhibit a few operators that belong to these representations.  In order to construct operators, we can use the fields in the Lagrangian, as well as monopole operators.  The monopole operators $M^{n_1,\dots, n_N}_{\tilde n_1,\dots, \tilde n_N}$  create $\diag\{n_1, \ldots, n_N\}$ and $\diag\{\tilde n_1, \ldots, \tilde n_N\}$ units of magnetic flux through the two gauge groups, respectively.  Here, we take both the $n_r$ and $\tilde n_r$ to be in descending order.  If the gauge groups are $U(N)$, then the equations of motion imply that
 \es{sumn}{
  \sum_r n_r = \sum_r \tilde n_r = - 2 Q_T \,,
 }
where $Q_T$ is the charge under the $U(1)_T$ symmetry mentioned above, quantized in half-integer units.\footnote{The conserved current associated to this charge is $
J^{\mu}=-\frac{1}{16\pi}\epsilon^{\mu\nu\rho}\left(\Tr F_{\nu\rho}+\Tr\tilde{F}_{\nu\rho}\right)
$.  The other linear combination of gauge field strengths vanishes by the equations of motion.}  If the gauge groups are $SU(N)$, then the $n_r$ and $\tilde n_r$ must each sum to zero.   We will only be considering BPS monopole operators, with zero R-charge.  In general, the R-charge is
\es{monoDim}{
E=\sum_{r,s=1}^N \left[ |n_r-\tilde n_s|- \frac 12 |n_r-n_s|- \frac 12 |\tilde n_r-\tilde n_s| \right]\,,
}
as was first proposed in \cite{Gaiotto:2008ak} and derived in \cite{Kim:2009wb, Benna:2009xd, Bashkirov:2010kz}.  It is easy to see from \eqref{monoDim} that $E=0$ only for $n_r=\tilde n_r$. In order to avoid clutter, we denote such operators simply by $M^{n_1,\dots, n_N}$.  For $k\neq0$ these monopole operators transform nontrivially under the gauge group in a way to be described shortly.  To form gauge-invariant operators, the monopole operators $M^{n_1,\dots, n_N}$ need to be dressed with the matter fields.  Let us show this explicitly for the lowest few multiplets.

\subsubsection{ABJM$_{3, 1}$}

For the ABJM$_{3, 1}$ theory, the monopole operators $M^{n_1, n_2, n_3}$ transform under the $U(3) \times U(3)$ gauge group as
\es{ABJMirrep}{
&U(3)\times U(3) \text{    irrep:}\quad\left(\Upsilon_\nu,\overline\Upsilon_{-\nu}\right)\,,\quad
  \Upsilon=\,\, 
  \raisebox{ -3ex}{$\overset{  \overbrace{ \begin{ytableau}
 {} &{}&{}\\
 {} &{}\\
\end{ytableau}  }^{n_1-n_3}}{\underbrace{\hphantom{aaaaaa}}_{n_2-n_3} \hphantom{aa\,\,}}  $ } \,,\quad \nu=\sum_r n_r\,,
}
where we have denoted a $U(3)$ irrep by $\Upsilon_\nu$, where $\Upsilon$ is an $SU(3)$ Young diagram and $\nu$ is the charge under the diagonal $U(1)$, normalized such that the fundamental of $U(3)$ is $\square_1$.  In \eqref{ABJMirrep}, $\overline\Upsilon$ denotes the conjugate tableau to $\Upsilon$.  In particular, we can write $M^{n_1, n_2, n_3}$ more explicitly as a symmetric traceless tensor with $n_1 - n_2$ fundamental and $n_2 -n_3$ anti-fundamental indices under the first gauge group, and $n_1 - n_2$ anti-fundamental and $n_2 -n_3$ fundamental indices under the second gauge group.  Using a notation in which $U(3)$ fundamental indices are upper and anti-fundamental indices are lower, this is  $(M^{n_1, n_2, n_3})^{\alpha_1 \ldots \alpha_{n_1 - n_2} \dot \beta_1 \ldots \dot \beta_{n_2 -n_3}}_{\beta_1 \ldots \beta_{n_2 -n_3} \dot \alpha_1 \ldots \dot \alpha_{n_1 - n_2}}$.

We can construct gauge invariant BPS states with nonzero $Q_T$ by dressing $M^{n_1,n_2,n_3}$ with appropriate products of 
$C^I=(A_1,A_2,B_1^\dagger ,B_2^\dagger)$ and $C_I^\dagger$, where upper/lower $I=1,2,3,4$ is a fundamental/anti-fundamental index for $SU(4)_R$. In the notation above, the $C^I$ transform in the gauge irrep $(\Upsilon_\nu, \overline{\Upsilon}_{-\nu})$, with $\Upsilon= \square$ and $\nu = 1$.  Including explicit gauge indices, we would write $(C^I)^
\alpha_{\dot \alpha}$ and $(C_I^\dagger)^{\dot \alpha}_{\alpha}$.

Using a single matter field, we find that $C^\dagger_IM^{1,0,0}$ and $C^IM^{0,0,-1}$ (with the gauge indices contracted in the only possible way, namely $(C^\dagger_I)^{\dot \alpha}_\alpha (M^{1,0,0})^\alpha_{\dot \alpha}$ and $(C^I)^\alpha_{\dot \alpha} (M^{0,0,-1})^{\dot \alpha}_\alpha$) are the only gauge-invariant combinations.  They transform under $SU(4)_R\times U(1)_T$ as $\overline{\bold{4}}_{-\frac12}$ and ${\bold{4}}_{-\frac12}$, respectively. These operators have scaling dimension $\Delta = \frac12$, and are thus free.  They are part of the free sector of ABJM theory, which also contains all operators that appear in the OPE of  $C^\dagger_IM^{1,0,0}$ and $C^IM^{0,0,-1}$.  The lowest few scalar operators in this free sector are given schematically in Table~\ref{free}.\footnote{The relegation of operators to the free, mixed, and interacting sectors is schematic, as there may be mixing between operators in the same representations.}   The hallmark of the free sector is the $OSp(8|4)$ irrep $(B, +)^{[0010]}$ whose scalar operators were mentioned above.  Another feature is the presence of a stress tensor multiplet $(B, +)^{[0020]}$.
\begin{table}[!h]
\begin{center}
\begin{tabular}{c||c|c|c}
$\cO$ & $\Delta$ & $SU(4)_R\times U(1)_T$ & $OSp(8|4)$ irrep \\
 \hline \hline
     $ C^\dagger_{I}M^{1,0,0}$ & $\frac12 $&$ \overline{\bold {4}}_{-\frac12}$&$ \raisebox{-1ex}{$(B,+)^{[0010]}$}$\\
  \cline{1-3}
    $ C^{I}M^{0,0,-1}$ & $\frac12 $&$ {\bold {4}}_{\frac12}$&\\
  \hline
    $ C^\dagger_{(I}C^\dagger_{J)}M^{2,0,0}$ & $1 $&$ \overline{\bold {10}}_{-1}$&\\
  \cline{1-3}
   $ C^{(I}C^{J)}M^{0,0,-2}$ & $1 $&$ {\bold {10}}_{1}$&$(B,+)^{[0020]}$\\
 \cline{1-3}
   $ C^\dagger_{I}C^{J}M^{1,0,-1}$ & $1 $&$ {\bold {15}}_{0}$&\\
  \hline
  $ C^\dagger_{(I}C^\dagger_{J}C^\dagger_{K)}M^{3,0,0}$ & $\frac32 $&$ {\bold {20}}^{\prime\prime}_{-\frac32}$&\\
   \cline{1-3}
    $ C^{(I}C^{J}C^{K)}M^{0,0,-3}$ & $\frac32 $&$\overline {\bold {20}}^{\prime\prime}_{\frac32}$&$ \raisebox{-1ex}{$(B,+)^{[0030]}$}$\\
 \cline{1-3}
$ C^{I}C^\dagger_{(J}C^\dagger_{K)}M^{2,0,-1}$&$\frac32 $& $ \overline{\bold {36}}_{-\frac12}$&\\
 \cline{1-3}
 $ C^\dagger_{I}C^{(J}C^{K)}M^{1,0,-2}$& $\frac32 $&$ {\bold {36}}_{\frac12}$&\\
  \hline
  \end{tabular}
\caption{BPS operators with $\Delta\leq\frac32$ in the free sector of the ABJM$_{3, 1}$ theory. \label{free}}
\end{center}
\end{table}

The interacting sector, whose lowest few scalar operators are given schematically in Table~\ref{inter}, consists of all operators that decouple from the free sector.  For instance, the first operator in Table~\ref{inter}, $C^\dagger_{(I}C^\dagger_{J)}M^{1,1,0}$, can be written more explicitly as:
\es{example}{
\epsilon^{\alpha\beta\gamma} \epsilon_{\dot \alpha \dot \beta \dot \gamma} C^{\dagger}_{(I}{}^{[\dot \alpha}_{[\alpha}C^\dagger_{J)}{}^{\dot \beta]}_{\beta]}M^{\dot \gamma}_{\gamma}\,.
}
Note that the flavor indices are symmetrized, because the gauge indices for both gauge groups are simultaneously anti-symmetrized, and thus this operator transforms in the $\overline{\bf 10}$ of $SU(4)_R$ and has $U(1)_T$ charge $-1$.  Also note the presence of another stress tensor multiplet $(B, +)^{[0020]}$, which is different from the one appearing in the free sector.  Thus, this ABJM theory has two ${\cal N} = 8$ stress tensor multiplets, each corresponding to a decoupled sector. 

\begin{table}[!ht]
\begin{center}
\begin{tabular}{c||c|c|c}
$\cO$ & $\Delta$ & $SU(4)_R\times U(1)_T$ & $OSp(8|4)$ irrep \\
 \hline \hline
        $ C^\dagger_{(I}C^\dagger_{J)}M^{1,1,0}$ & $1 $&$ \overline{\bold {10}}_{-1}$&\\
  \cline{1-3}
   $ C^{(I}C^{J)}M^{0,-1,-1}$ & $1 $&$ {\bold {10}}_{1}$&$(B,+)^{[0020]}$\\
 \cline{1-3}
   $ C^\dagger_{I}C^{J}$ & $1 $&$ {\bold {15}}_{0}$&\\
  \hline
  $ C^\dagger_{(I}C^\dagger_{J}C^\dagger_{K)}M^{1,1,1}$ & $\frac32 $&$ {\bold {20}}^{\prime\prime}_{-\frac32}$&\\
   \cline{1-3}
    $ C^{(I}C^{J}C^{K)}M^{-1,-1,-1}$ & $\frac32 $&$\overline {\bold {20}}^{\prime\prime}_{\frac32}$&$ \raisebox{-1ex}{$(B,+)^{[0030]}$}$\\
 \cline{1-3}
$ C^{I}C^\dagger_{(J}C^\dagger_{K)}M^{1,0,0}$&$\frac32 $& $ \overline{\bold {36}}_{-\frac12}$&\\
 \cline{1-3}
 $ C^\dagger_{I}C^{(J}C^{K)}M^{0,0,-1}$& $\frac32 $&$ {\bold {36}}_{\frac12}$&\\
  \hline
  \end{tabular}
\caption{BPS operators with $\Delta\leq\frac32$ in the interacting sector of the ABJM$_{3, 1}$ theory. \label{inter}}
\end{center}
\end{table}

Lastly, there is a mixed sector whose lowest few scalar operators are given in Table~\ref{mix},\footnote{The appearance of $C^I C^\dagger_{(J}C^\dagger_{K)}M^{1,0,0}$ in both the mixed and interacting sector is because there are two singlets in the product $\bold3\otimes\bold3\otimes\bar{\bold3}\otimes\bar{\bold3}$ of gauge irreps, and thus two inequivalent ways of contracting the gauge indices.} which consists of all operators built using both free and interacting sector operators. Note that there are no free or stress tensor multiplets in the mixed sector, as expected, but there are now both $(B,+)$ and $(B,2)$ operators with dimension $\frac32$. \\
\begin{table}[!h]
\begin{center}
\begin{tabular}{c||c|c|c}
$\cO$ & $\Delta$ & $SU(4)_R\times U(1)_T$ & $OSp(8|4)$ irrep \\
 \hline \hline
    $ C^\dagger_{(I}C^\dagger_{J}C^\dagger_{K)}M^{2,1,0}$ & $\frac32 $&$ {\bold {20}}^{\prime\prime}_{-\frac32}$&\\
   \cline{1-3}
    $ C^{(I}C^{J}C^{K)}M^{0,-1,-2}$ & $\frac32 $&$ \overline{\bold {20}}^{\prime\prime}_{\frac32}$&$ \raisebox{-1ex}{$(B,+)^{[0030]}$}$\\
 \cline{1-3}
$ C^{I}C^\dagger_{(J}C^\dagger_{K)}M^{1,1,-1}$&$\frac32 $& $ \overline{\bold {36}}_{-\frac12}$&\\
 \cline{1-3}
 $ C^\dagger_{I}C^{(J}C^{K)}M^{1,-1,-1}$& $\frac32 $&$ {\bold {36}}_{\frac12}$&\\
  \hline
     $ C^\dagger_{(I}C^\dagger_{[J)}C^\dagger_{K]}M^{2,1,0}$ & $\frac32 $&$ {\bold {20}}_{-\frac32}$&\\
   \cline{1-3}
    $ C^{(I}C^{[J)}C^{K]}M^{0,-1,-2}$ & $\frac32 $&$ \overline{\bold {20}}_{\frac32}$&\\
 \cline{1-3}
$ C^{I}C^\dagger_{(J}C^\dagger_{K)}M^{1,0,0}$&$\frac32 $& $ \overline{\bold {36}}_{-\frac12}$&\\
 \cline{1-3}
 $ C^\dagger_{I}C^{(J}C^{K)}M^{0,0,-1}$& $\frac32 $&$ {\bold {36}}_{\frac12}$&$ \raisebox{-1ex}{$(B,2)^{[0110]}$}$\\
  \cline{1-3}
$ C^{I}C^\dagger_{[J}C^\dagger_{K]}M^{1,0,0}$&$\frac32 $& $ \overline{\bold {20}}_{-\frac12}$&\\
 \cline{1-3}
 $ C^\dagger_{I}C^{[J}C^{K]}M^{0,0,-1}$& $\frac32 $&$ {\bold {20}}_{\frac12}$&\\
 \cline{1-3}
$ C^{I}C^\dagger_{[I}C^\dagger_{J]}M^{1,0,0}$&$\frac32 $& $ \overline{\bold {4}}_{-\frac12}$&\\
 \cline{1-3}
 $ C^\dagger_{I}C^{[I}C^{J]}M^{0,0,-1}$& $\frac32 $&$ {\bold {4}}_{\frac12}$&\\
  \hline
  \end{tabular}
\caption{BPS operators with $\Delta\leq\frac32$ in the mixed sector of ABJM$_{3, 1}$ theory. \label{mix}}
\end{center}
\end{table}

\subsubsection{BLG$_3$}

A similar construction holds for the BLG$_3$ theory.  One difference between this theory and the ABJM$_{3, 1}$ example we studied above is that the BLG$_3$ theory has a different set of monopole operators with $E=0$, labeled by only a single positive half-integer GNO charge $n$.  They transform in the $SU(2) \times SU(2)$ gauge irrep
\es{BLGDynk}{
SU(2)\times SU(2)\text{    irrep:}\qquad \left({\bf 6n+1},{\bf 6n+1}\right)\,.
}
(For the BLG$_k$ theory with arbitrary $k$, the gauge irrep is $({\bf 2kn+1},{\bf 2kn+1})$.)  These monopole operators must be combined with the matter fields $C^I$ and $C^\dagger_I$, each of which transform as $({\bf 2}, {\bf 2})$ under the gauge group.

The lowest dimension gauge invariant operators are quadratic in $C^I$ and $C^\dagger_I$ and do not require monopole operators.  The next lowest are cubic in the $C^I$ and $C^\dagger_I$ and require the monopole operator with $n=1/2$.  See Table~\ref{BLG}.  These operators are in one-to-one correspondence with operators from the interacting sector of the ABJM$_{3, 1}$ theory given in Table \ref{inter}.  We take this match to be the first piece of evidence for the duality \eqref{NewDuality}--\eqref{NewDualityRephrase} between the two theories. \\

 \begin{table}[!h]
\begin{center}
\begin{tabular}{c||c|c|c}
$\cO$ & $\Delta$ & $SU(4)_R\times U(1)_t$ & $OSp(8|4)$ irrep \\
 \hline \hline
        $ C^\dagger_{(I}C^\dagger_{J)}$ & $1 $&$ \overline{\bold {10}}_{-1}$&\\
  \cline{1-3}
   $ C^{(I}C^{J)}$ & $1 $&$ {\bold {10}}_{1}$&$(B,+)^{[0020]}$\\
 \cline{1-3}
   $ C^\dagger_{I}C^{J}$ & $1 $&$ {\bold {15}}_{0}$&\\
  \hline
  $ C^\dagger_{(I}C^\dagger_{J}C^\dagger_{K)}M^{1/2}$ & $\frac32 $&$ {\bold {20}}^{\prime\prime}_{-\frac32}$&\\
   \cline{1-3}
    $ C^{(I}C^{J}C^{K)}M^{1/2}$ & $\frac32 $&$\overline {\bold {20}}^{\prime\prime}_{\frac32}$&$ \raisebox{-1ex}{$(B,+)^{[0030]}$}$\\
 \cline{1-3}
$ C^{I}C^\dagger_{(J}C^\dagger_{K)}M^{1/2}$&$\frac32 $& $ \overline{\bold {36}}_{-\frac12}$&\\
 \cline{1-3}
 $ C^\dagger_{I}C^{(J}C^{K)}M^{1/2}$& $\frac32 $&$ {\bold {36}}_{\frac12}$&\\
  \hline
  \end{tabular}
\caption{BPS operators with $\Delta\leq\frac32$ in the BLG$_3$ theory. \label{BLG}}
\end{center}
\end{table}

\subsection{Superconformal Index}

As an alternative to the explicit construction given in the previous section, one can use the superconformal index.    The superconformal index, to be defined more precisely shortly, captures information about protected representations of the superconformal algebra.  Its advantage over the explicit construction of the previous section is that it can be rigorously computed using supersymmetric localization.  Its disadvantage is that the information it encodes does not unambiguously identify all the $\mathfrak{osp}(8|4)$ representations.  

In order to define the superconformal index, it is convenient to view an ${\cal N} = 8$ SCFT as an ${\cal N} = 2$ SCFT with $SU(4)$ flavor symmetry.  One can then consider a supercharge $Q$ within the ${\cal N} = 2$ superconformal algebra such that $\{Q, Q^\dagger\} = \Delta - R - j_3$, where $\Delta$ is the scaling dimension, $j_3$ is the third component of the angular momentum, and $R$ is the $U(1)_R$ charge.  (There is a unique such supercharge, and it has $\Delta = 1/2$, $R = 1$, and $j_3 = -1/2$.)  The superconformal index with respect to $Q$ is defined as the trace over the $S^2 \times \R$ Hilbert space
\es{index}{
I(x,z_f)=\Tr\left[(-1)^Fx^{\Delta+j_3}\prod_{f=1}^3 z_f^{F_f}\right] \,,
}
where $F = (-1)^{2j_3}$ is the fermion number and $F_f$ are the charges under the Cartan of the $SU(4)$ flavor symmetry.  Standard arguments imply that the only states contributing to the trace in \eqref{index} obey $\Delta = R + j_3$;  all others cancel pairwise.

The indices for the theories we are interested in have been computed using supersymmetric localization in \cite{Honda:2012ik}, following the general computation in \cite{Kim:2009wb}.  It can be shown that $I_\text{ABJM$_{3, 1}$} = I_\text{BLG$_3$}  I_\text{free}$, where $I_\text{ABJM$_{3, 1}$}$ is the index of the ABJM$_{3, 1}$ theory, $I_\text{BLG$_3$}$ is that of the BLG$_3$ theory, and $I_\text{free}$ is that of the ABJM$_{1, 1}$ theory, which is free.  For instance, keeping only one fugacity $z$ corresponding to the Cartan element of $SU(4)$ given by either $U(1)_T$ or $U(1)_t$, we have\footnote{We fix a typo in \cite{Honda:2012ik} for the coefficient of $z^2x^3$ in the expression for $I_\text{ABJM$_{3, 1}$}$.} 
\es{Indices}{
I_\text{ABJM$_{3, 1}$} &= 1+8x+71x^2+320x^3+2z\left(x^{1/2}+12x^{3/2}+78x^{5/2}\right)\\
&+z^2\left(6x+56x^2+295x^3\right)+z^3(14x^{3/2}+114x^{5/2})+O(z^4,x^{7/2})+\left(z\leftrightarrow z^{-1}\right)\,, \\
I_\text{free} &=1+4x+x^2+4x^3+2z\left(x^{1/2}+2x^{3/2}\right)+z^2\left(3x+4x^2\right)\\
&+4z^3\left(x^{3/2}+x^{5/2}\right)+O(z^4,x^{7/2})+\left(z\leftrightarrow z^{-1}\right)\,, \\
I_\text{BLG$_3$} &=1+4x+12x^2+24x^3+2z\left(3x^{3/2}+11x^{5/2}\right)\\
&+z^2\left(3x+8x^2+27x^3\right)+2z^3(2x^{3/2}+10x^{5/2})+O(z^4,x^{7/2})+\left(z\leftrightarrow z^{-1}\right)\,.
}
One can indeed check that these expressions obey $I_\text{ABJM$_{3, 1}$} = I_\text{BLG$_3$}  I_\text{free}$ up to the order given.  We regard this match of the indices as the second piece of evidence supporting our conjectured duality \eqref{NewDuality}--\eqref{NewDualityRephrase}.

\section{Moduli Space}
\label{modSpace}
We now show how to relate the (classical) moduli space of vacua of the ABJM$_{3, 1}$ theory to that of the BLG$_3$ theory.  The moduli space can be found by modding out the zero locus of the scalar potential by the gauge transformations.  For both theories, one can check that the scalar potential vanishes provided that \cite{VanRaamsdonk:2008ft,Aharony:2008ug} 
\es{diag}{
\langle A_{a\beta}^{\dot \alpha}\rangle=a_{\beta a}\delta^{\dot \alpha}_\beta\,,\qquad \langle B_{\dot{a}\dot \alpha}^{ \beta}\rangle=b^{ \beta}_{\dot a}\delta^{\dot \alpha}_{\beta}\,,\qquad
}
where $a_{\beta a}, b^{ \beta}_{\dot a}$ are complex numbers, and where we used part of the gauge symmetry to put $A_{a\beta}^{\dot \alpha}$ and $B_{\dot{a}\dot \alpha}^{ \beta}$ in diagonal form. For a gauge group of rank $N$, the moduli space is thus parameterized by $4N$ complex numbers $z_r = \{ a_{r1}, a_{r2}, b^r_1, b^r_2 \}$ for $r=1,\dots,N$, modulo residual gauge transformations. 

The residual gauge symmetry gives further relations on $z_r$. For the ABJM$_{3, 1}$ theory, these relations are \cite{Aharony:2008ug}
\es{modABJM}{
z_r\sim z_{\sigma(r)}\,,\qquad \sigma\in S_3\,,
}
where $r=1,2,3$ and $S_3$ is the symmetric group of order six. The moduli space is thus $(\mathbb{C}^4)^3/S_3$. From the M-theory perspective, this is the moduli space of three M2-branes in flat space, where the $S_3$ corresponds to permuting the indistinguishable branes. 

For the BLG$_3$ theory, for which we denote the corresponding coordinates by $z_r'$ instead of $z_r$, the relations are \cite{Lambert:2008et,Lambert:2010ji,Distler:2008mk}
\es{modBLG}{
z'_1\sim z'_2\,,\qquad z'_1\sim e^{2\pi i/3} z'_1\,,\qquad z'_2\sim e^{-2\pi i/3}z'_2\,.
}
The first relation comes from permuting the identical gauge groups, while the last two come from identifications that depend on the Chern-Simons coupling. These relations define the moduli space $(\mathbb{C}^4)^2/\bold{D}_3$, where $\bold{D}_3$ is the dihedral group of order six. We wish to identify this with the interacting sector of ABJM$_{3, 1}$. To distinguish between the free and interacting sector of the latter, consider the  reparameterization
\es{modABJM2}{
w_1=e^{-2\pi i/3}z_1+e^{2\pi i/3}z_2+z_3\,,\qquad w_2=e^{2\pi i/3}z_1+e^{-2\pi i/3}z_2+z_3\,,\qquad w_3=z_1+z_2+z_3\,.
}
The parameter $w_3$ is invariant under $S_3$ and thus parameterizes the moduli space of the free theory. The interacting sector is parameterized by $w_1,w_2$, which transform under the permutations $(12),(123)\in S_3$ as
\es{dihedral}{
(12):\qquad w_1&\sim w_2\,,\\
(123):\qquad w_1&\sim e^{2\pi i/3} w_1\,,\qquad w_2\sim e^{-2\pi i/3}w_2\,,
}
where $(12)$ permutes $z_1\leftrightarrow z_2$ and $(123)$ permutes $z_1\to z_2\,,z_2\to z_3\,,z_3\to z_1$. These relations are the same as \eqref{modBLG}, which establishes the isomorphism
\es{simToDi}{
(\mathbb{C}^4)^3/S_3\cong  (\mathbb{C}^4)^2/\bold{D}_3\times \mathbb{C}^4\,,
}
where $\mathbb{C}^4$ corresponds to the free sector of the ABJM$_{3, 1}$ theory, and $(\mathbb{C}^4)^2/\bold{D}_3$ corresponds to the interacting sector as well as to the BLG$_3$ theory.  We regard the match between the moduli spaces \eqref{simToDi} as the third piece of evidence supporting our conjectured duality \eqref{NewDuality}--\eqref{NewDualityRephrase}.

\section{The $S^3$ Partition Function}
\label{S3partition}

We will now compare the $S^3$ partition functions of the two theories. The partition function for the ABJM$_{N, k}$ theory can be written as the following finite dimensional integral \cite{Kapustin:2009kz}:
\es{partitionABJM}{
Z _{\text{ABJM}_{N, k}}= \frac{1}{(N!)^2} \int d^N \sigma d^N \widetilde{\sigma} \:
e^{\pi i k \sum_{\alpha=1}^N (\sigma_\alpha^2 - \widetilde{\sigma}_\alpha^2)} \:
\left(\frac{\prod_{\alpha<\beta} 2 \sinh(\pi(\sigma_\alpha - \sigma_\beta))  2 \sinh(\pi ({\widetilde \sigma}_\alpha - {\widetilde \sigma}_\beta))}{\prod_{\alpha,\beta} 2 \cosh(\pi (\sigma_\alpha - \widetilde{\sigma}_\beta ))}\right)^2 \,,
}
where $\sigma_\alpha,\widetilde\sigma_\alpha$ are integration variables that can be interpreted as the eigenvalues of the scalars in the vector multiplets associated with the two $U(N)$ gauge groups.
For $k=1$ and $N=1,3$ we find
\es{partitionABJM3}{
Z _\text{ABJM$_{3, 1}$}= \frac{\pi - 3}{64\pi}\,,\qquad  Z _\text{ABJM$_{1, 1}$}=Z _\text{free}= \frac{1}{4}\,,
}
where recall that the ABJM$_{1, 1}$ theory is free. 

The partition function of the BLG$_k$ theory can be derived from the ABJM$_{N, k}$  partition function \eqref{partitionABJM} by setting $N=2$, imposing the constraints $\sigma_1+\sigma_2=\widetilde\sigma_1+\widetilde\sigma_2=0$, and multiplying by $2$ to take into account the $\Z_2$ quotient in the $(SU(2) \times SU(2)) / \Z_2$ gauge group. The result is
\es{partitionBLGVal}{
Z _\text{BLG$_k$}= \frac{1}{64 \pi^2} \int d^2\sigma_\pm \: e^{\frac{2k i \sigma_+ \sigma_-}{\pi}} \left(
\frac{\sinh(\sigma_+ + \sigma_-)\sinh(\sigma_+ - \sigma_-)}{\cosh^2(\sigma_+) \cosh^2(\sigma_-)}
 \right)^2 \,,
 }
 where we have made the change of variables $\sigma_\pm = \pi(\sigma_1 \pm \widetilde{\sigma}_1)$. For $k=3$, we find that 
 \es{partitionFinal}{
 Z _\text{BLG$_3$}= \frac{\pi - 3}{16\pi}=\frac{Z _\text{ABJM$_{3, 1}$} }{Z_\text{free}} \,,
}
as we expect from our duality.  We regard \eqref{partitionFinal} as our fourth piece of evidence supporting the conjectured duality \eqref{NewDuality}--\eqref{NewDualityRephrase}.

\section{One-Dimensional Topological Sector}
\label{topStuff}

Lastly, let us attempt to make a more detailed check of the duality \eqref{NewDuality}--\eqref{NewDualityRephrase} at the level of correlation functions of BPS operators.  As explained in \cite{Chester:2014mea,Beem:2016cbd,Beem:2013sza}, abstract arguments based on the superconformal algebra show that all three-dimensional ${\cal N} \geq 4$ SCFTs have two one-dimensional topological sectors (defined either on a line in flat space or on a great circle within $S^3$), one associated with the Higgs branch and the other with the Coulomb branch.  More precisely, these topological sectors arise as follows.  All ${\cal N} = 4$ SCFTs have an $SU(2)_H \times SU(2)_C$ R-symmetry.  In general, there can be two types of $1/2$-BPS scalar operators in these theories:   ``Higgs branch operators'' that are invariant under $SU(2)_C$ and have scaling dimension $\Delta$ equal to the $SU(2)_H$ spin $j_H$,  and ``Coulomb branch operators'' that are invariant under $SU(2)_H$ and have scaling dimension $\Delta$ equal to the $SU(2)_C$ spin $j_C$.  The operators belonging to the Higgs (Coulomb) branch topological sector are linear combinations of the first (second) class of $1/2$-BPS operators above with specific position-dependent coefficients.  These operators, when inserted on a line in flat space or on a great circle on $S^3$, have topological correlation functions because they represent non-trivial cohomology classes of a nilpotent supercharge with respect to which translations along the line / circle are exact.  Concretely, in the case where the 1d Higgs branch theory is defined on a great circle parameterized by $\vphi \in [-\pi/2, \pi/2)$ that sits within a round $S^3$ of radius $r$, the 1d operators are
 \es{Ophi}{
  {\cal O}(\vphi) = {\cal O}_{i_1 \ldots i_{2j_H}}(\vphi) u^{i_1}(\vphi) \ldots u^{i_{2j_H}}(\vphi)  \,, \qquad
   u^i(\vphi) \equiv \begin{pmatrix}
    \cos (\vphi/2) \\
    \sin (\vphi/2) 
   \end{pmatrix} \,,
 }
where ${\cal O}_{i_1 \ldots i_{2j_H}}(\vphi)$ is a 3d operator with $\Delta = j_H$ and $j_C = 0$, written as a symmetric, rank-$2j_H$ tensor of $SU(2)_H$. For more details, see~\cite{Dedushenko:2016jxl} as well as \cite{Chester:2014mea,Beem:2016cbd,Beem:2013sza}.

 For the particular case of  ${\cal N} = 8$ SCFTs, the Higgs and Coulomb topological sectors are isomorphic, so without loss of generality we will study the Higgs one.  In  \cite{Dedushenko:2016jxl}, it was shown that for ${\cal N} = 4$ SCFTs described by a Lagrangian with a vector multiplet with gauge algebra $\mathfrak{g}$ and a hypermultiplet in representation ${\cal R}$ of $\mathfrak{g}$, it is possible to use supersymmetric localization to obtain an explicit description of the 1d sector associated with the Higgs branch.  When the 1d topological sector is defined on a great circle within $S^3$ parameterized by $\vphi$, as above,  its explicit description takes the form of a Gaussian 1d theory coupled to a matrix model:
  \es{1dSector}{
   Z = \int_{\text{Cartan of $\mathfrak{g}$}} d\sigma\, \text{det}{}_\text{adj}^\prime (2 \sinh(\pi \sigma)) \int DQ\, D\tQ\, e^{4 \pi r \int d\vphi\, \left(  \tQ \partial_\vphi Q + \tQ \sigma Q \right)} \,.
  }
Here, $\sigma$ is the matrix degree of freedom that has its origin in the 3d vector multiplet and was diagonalized to lie within the Cartan of the gauge algebra.  The 1d fields $Q(\vphi)$ and $\tQ(\vphi)$ have their origin in the 3d hypermultiplet and transform in the representations ${\cal R}$ and $\overline{{\cal R}}$, respectively. Their definition in terms of the hypermultiplet scalars is as in \eqref{Ophi}, with ${\cal O}$ replaced by the hypermultiplet scalars transforming in the fundamental of $SU(2)_H$. Upon integrating out $Q$ and $\tQ$ one obtains the Kapustin-Willett-Yaakov matrix model \cite{Kapustin:2009kz} for the $S^3$ partition function of the ${\cal N} = 4$ SCFT:
 \es{KWY}{
  Z = \int_{\text{Cartan of $\mathfrak{g}$}} d\sigma\,
   \frac{\text{det}{}_\text{adj}^\prime (2 \sinh(\pi \sigma)) }{\text{det}_{\cal R} (2 \cosh(\pi \sigma)) } \,.
 } 
The description \eqref{1dSector} can be used to calculate arbitrary $n$-point functions of operators belonging to the 1d sector, so this result opens up the possibility of performing more detailed tests of our proposed duality \eqref{NewDuality}--\eqref{NewDualityRephrase} involving correlation functions captured by the 1d sector.  Unfortunately, the ABJM and BLG theories we are interested in do not have Lagrangian descriptions in terms of just vector multiplets and hypermultiplets (one cannot accommodate non-zero Chern-Simons levels with just vector multiplets and hypermultiplets), so the result \eqref{1dSector} quoted above does not directly apply to these theories.
 
Fortunately, there is a way around this difficulty.  The right-hand side of \eqref{NewDuality}--\eqref{NewDualityRephrase}, or more generally the ABJM$_{N, 1}$ theory, has a dual description as an ${\cal N} = 4$ $U(N)$ gauge theory coupled to an adjoint hypermultiplet and a fundamental hypermultiplet \cite{deBoer:1996ck, Aharony:2008ug, Bashkirov:2011pt}. So if we worked with this dual description we could use \eqref{1dSector} to compute correlation functions in the Higgs branch topological sector, and we will do so in the case of interest $N=3$.   For the BLG theories no such dual description is available, but we will conjecture that a modification of \eqref{1dSector} will allow us to compute some of the correlation functions in the Higgs branch sector.  Our conjecture is that to the integrand of \eqref{1dSector} we should insert 
 \es{insertion}{
  e^{i \pi k \tr \sigma^2}
 } 
for every gauge group factor that has a Chern-Simons level $k$, where the trace is taken in the fundamental representation of that gauge group factor and in the trivial representation of the rest.  This conjecture is motivated by the fact that this is the correct prescription in the matrix model \eqref{KWY}.  Importantly, it allows us to compute correlation functions of gauge-invariant operators built from $Q$ and $\tQ$.  However, unlike when $k=0$, these operators are not the most general operators in the 1d theory;  some of the operators in the 1d theory descend from 3d monopole operators, and these are not captured by \eqref{1dSector} supplemented by \eqref{insertion}.   Nevertheless, we will still be able to compute correlation functions of non-monopole operators in the BLG$_3$ theory and compare them with the analogous correlators in the ABJM${}_{3,1}$ theory.  As we will see, the results of these computations are consistent with our proposed duality in \eqref{NewDuality}--\eqref{NewDualityRephrase}.

From the $\mathcal{N}=8$ perspective, the operators in the Higgs branch topological theory are specific linear combinations of at least $1/4$-BPS short representations \cite{Chester:2014mea}.  To be concrete, let us consider an $SU(2)_H \times SU(2)_C \times SU(2)_F \times SU(2)_{F'}$ subgroup of the ${\cal N} = 8$ $SO(8)_R$ R-symmetry, where, from the ${\cal N} = 4$ point of view, $SU(2)_H \times SU(2)_C$ is interpreted as the R-symmetry and $SU(2)_F \times SU(2)_{F'}$ as a flavor symmetry.  One can consider this embedding such that the fundamental representations of $SO(8)$ have the following decompositions:
 \es{FundDecomp}{
  [1000] = {\bf 8}_v &\to ({\bf 2}, {\bf 2}, {\bf 1}, {\bf 1} ) \oplus ({\bf 1}, {\bf 1}, {\bf 2}, {\bf 2} ) \,, \\
  [0010] = {\bf 8}_c &\to ({\bf 2}, {\bf 1}, {\bf 2}, {\bf 1} ) \oplus ({\bf 1}, {\bf 2}, {\bf 1}, {\bf 2} ) \,, \\
  [0001] = {\bf 8}_s &\to ({\bf 2}, {\bf 1}, {\bf 1}, {\bf 2} ) \oplus ({\bf 1}, {\bf 2}, {\bf 2}, {\bf 1} ) \,.
 }
A careful analysis \cite{Chester:2014mea} shows that the only operators in the 1d theory come from the superconformal primaries of ${\cal N} = 8$ multiplets that are at least $1/4$-BPS---in our case, these will be the $(B, +)^{[00m0]}$ and $(B, 2)^{[0nm0]}$ representations.  The superconformal primaries of these multiplets are scalars with scaling dimension $\Delta = n+m/2$ and $SO(8)_R$ irrep with Dynkin labels $[0nm0]$.  They give 1d operators that are singlets of $SU(2)_{F'}$ and that transform in the spin-$m/2$ representation of  $SU(2)_F$:  
 \es{Representatives}{
  (B, +)^{[00m0]}&: \qquad {\cal O}^{(\Delta, j_F)}_{a_1 \ldots a_{2 j_F}}(\vphi) \qquad 
   \Delta = j_F = \frac{m}{2} \,, \\
   (B, 2)^{[0 n m 0]}&: \qquad {\cal O}^{(\Delta, j_F)}_{a_1 \ldots a_{2 j_F}}(\vphi)  \qquad 
   \Delta = j_F + n = \frac{m}{2} + n \,, 
 } 
where we have denoted the 1d operators as ${\cal O}^{(\Delta, j_F)}_{a_1 \ldots a_{2 j_F}}(\vphi)$, writing them explicitly as rank-$2j_F$ symmetric tensors of the $SU(2)_F$.   This $SU(2)_F$ is thus a global symmetry of the 1d topological theory.  

As in \cite{Chester:2014mea}, in order keep track of the $SU(2)_F$ indices more efficiently, we introduce polarization variables $ y^{ a}$, $ a=1,2$, and denote the operators in the 1d theory by 
\es{twistOp}{
\cO^{(\Delta, j)}(\varphi, y)=\cO^{(\Delta,j)}_{a_1 \ldots a_{2j}}(\varphi,y) y^{a_1} \cdots y^{a_{2j}}\,,
}
where in order to avoid clutter we simply denote $j_F=j$.   We consider a basis of 1d operators with diagonal two-point functions, normalized such that 
\es{twistNorm}{
  \braket{\cO^{(\Delta, j)}(\varphi_1,y_1) \cO^{(\Delta, j)}(\varphi_2,y_2)} &= \braket{y_1,y_2}^{2j} (\sgn  \varphi_{21})^{2\Delta} \,, \\
   \braket{\cO^{(\Delta_1, j_1)}(\varphi_1,y_1) \cO^{(\Delta_2, j_2)}(\varphi_2,y_2)} &= 0 \qquad
    \text{if $\cO^{(\Delta_1, j_1)} \neq \cO^{(\Delta_2, j_2)}$} \,,
}
where $\vphi_{21} \equiv \vphi_2 - \vphi_1$, and the product between $SU(2)_F$ polarizations is defined as
\es{su2-polarization-product}{
\braket{y_1, y_2} \equiv \epsilon_{ab} y_1^a y_2^b\,,  \qquad (\epsilon^{12} \equiv -\epsilon_{12} \equiv 1)\,.
}
The form of the three point functions is fixed by the $SU(2)_F$ symmetry up to an overall coefficient that we denote by $\lambda_{(\Delta_1,j_1),(\Delta_2,j_2),(\Delta_3,j_3)}$:
\es{twistNorm3}{
&\braket{\cO^{(\Delta_1, j_1)}(\varphi_1,y_1) \cO^{(\Delta_2, j_2)}(\varphi_2,y_2) \cO^{(\Delta_3, j_3)}(\varphi_3,y_3)} =\lambda_{(\Delta_1,j_1),(\Delta_2,j_2),(\Delta_3,j_3)}\\ 
&\qquad\qquad\times \braket{y_1,y_2}^{j_{123}} \braket{y_2,y_3}^{j_{231}} \braket{y_3,y_1}^{j_{312}} (\text{sgn} \: \varphi_{21})^{\Delta_{123}}(\text{sgn} \: \varphi_{32})^{\Delta_{231}}(\text{sgn} \: \varphi_{13})^{\Delta_{312}} \,,
}
where $j_{k_1 k_2 k_3} \equiv j_{k_1} + j_{k_2} - j_{k_3}$.  Eq.~\eqref{twistNorm3} is correct as long as $j_1$, $j_2$, and $j_3$ obey the triangle inequality.  If this requirement is not fulfilled, the RHS of \eqref{twistNorm3} vanishes.

\subsection{ABJM$_{3, 1}$}
\label{1dABJM}

Let us now apply the formalism introduced above to the $U(3)_1\times U(3)_{-1}$ ABJM theory in its dual description as a $U(3)$ gauge theory with both an adjoint and fundamental $\mathcal{N}=4$ hypermultiplet. The result \eqref{1dSector} reads in this case
\es{ABJM1d}{
Z_\text{ABJM$_{3, 1}$} &= \frac{1}{3!} \int d^3\sigma  \prod_{\alpha<\beta} 4 \sinh^2(\pi \sigma_{\alpha\beta})
\int {\cal D}Q^\alpha{\cal D}\widetilde{Q}_\alpha \int {\cal D}X^{\alpha}_{\:\: \beta} {\cal D}\widetilde{X}^{\:\: \beta}_{\alpha} e^{-S_\text{ABJM$_{3, 1}$}}\\
}
with
\es{ABJMact}{
S_\text{ABJM$_{3, 1}$} &= -4\pi r\int_{-\pi}^{\pi} d\varphi \left[\widetilde{Q}_\alpha \dot{Q}^\alpha+ \widetilde{X}^{\:\: \beta}_{\alpha} \dot{X}^{\alpha}_{\:\: \beta}
+ \sigma_\alpha \widetilde{Q}_\alpha Q^\alpha+ \sigma_{12}(\widetilde{X}^{\:\: 2}_{1}X^{1}_{\:\: 2} - \widetilde{X}^{\:\: 1}_{2}X^{2}_{\:\: 1}) \right. \\
&+ \left.  \sigma_{23}(\widetilde{X}^{\:\: 3}_{2}X^{2}_{\:\: 3} - \widetilde{X}^{\:\: 2}_{3}X^{3}_{\:\: 2}) +
\sigma_{31}(\widetilde{X}^{\:\: 1}_{3}X^{3}_{\:\:1} - \widetilde{X}^{\:\: 3}_{1}X^{1}_{\:\: 3}) \right] \,,
}
where $\alpha,\beta = 1,2,3$. The 1d fields $X^{\alpha}_{\:\: \beta}$ and $\widetilde{X}^{\:\: \beta}_{\alpha}$ correspond to the adjoint hypermultiplet, $Q^\alpha$ and $\widetilde{Q}_\alpha$ correspond to the fundamental hypermultiplet, and $\sigma_\alpha$ are the matrix degrees of freedom in the Cartan of the $U(3)$. 

The $D$-term relations of the 3d theory allow us to rewrite the $Q$'s in terms of the $X$'s, so we will only use the latter to construct operators.  Correlation functions of such operators can be computed by performing Wick contractions at fixed $\sigma$ with the propagator 
\es{Xprop}{
\braket{X^\alpha_{\:\: \beta}(\varphi_1,y_1)\widetilde{X}_\gamma^{\:\: \delta} (\varphi_2,y_2)}_\sigma = -\delta^\alpha_{\:\:\gamma} \delta^\delta_{\:\:\beta}\frac{\text{sgn} \: \varphi_{12} + \tanh(\pi \sigma_{\alpha\beta})}{8 \pi r} e^{- \sigma_{\alpha\beta} \varphi_{12}}\,.
}
and then integrating over $\sigma$:
\es{ABJMGen}{
\braket{\cO_1(\varphi_1,y_1) \cdots \cO_n(\varphi_n,y_n)} &= \frac{1}{Z _\text{ABJM$_{3,1}$}} \int d^3\sigma \: Z _\text{ABJM$_{3, 1}$}^\sigma \braket{\cO_1(\varphi_1,y_1) \cdots \cO_n(\varphi_n,y_n)}_\sigma\,,\\
&Z _\text{ABJM$_{3, 1}$}^\sigma = \frac{1}{2^6 \cdot 3!} \frac{\tanh^2(\pi \sigma_{12})\tanh^2(\pi \sigma_{13})\tanh^2(\pi \sigma_{23})} {\cosh(\pi \sigma_1) \cosh(\pi \sigma_2) \cosh(\pi \sigma_3)} \,,
}
where $\braket{\cdots}_\sigma$ is the correlation function for the Gaussian theory in \eqref{ABJMact} at fixed $\sigma$ computed using \eqref{ABJMGen}.

Being a 1d sector of an ${\cal N} = 8$ SCFT, the theory \eqref{ABJM1d} must have a flavor $SU(2)_F$ symmetry.  Indeed, it is not hard to see that the fields $(\widetilde{X}, X^T)$ transform as a doublet under $SU(2)_F$.  It is thus convenient to define 
\es{compositeA}{
{\cal X} (\varphi,y) = y^1 \widetilde{X}(\varphi,y) + y^2 X^T(\varphi,y)\,,
} 
where the $y^a$ are the same polarization variables introduced earlier in \eqref{twistOp}.

\subsubsection{Free Sector}

As explained above, the ABJM$_{3, 1}$ theory has a decoupled free sector.  Consequently, the 1d theory \eqref{ABJM1d} also has a decoupled free sector.  It is generated by the gauge invariant operator 
\es{freeMul}{
\cO^{(\frac12, \frac12)}_{\text{free}}(\varphi,y)=\tr\mathcal{X}(\varphi,y)\,,
}
which has its origin in the free multiplet $(B,+)^{[0010]}$, whose superconformal primaries are scalars of scaling dimension $\Delta = 1/2$. 

Since $\tr X$ and $\tr\widetilde X$ only appear in the kinetic term of \eqref{ABJMact}, we can simply read off the propagator
\es{freeProp}{
\langle\cO^{(\frac12,\frac12)}_{\text{free}}(\varphi_1, y_1)\cO^{(\frac12,\frac12)}_{\text{free}}(\varphi_2, y_2)\rangle= \frac{3}{8 \pi r}\braket{y_1,y_2}\sgn \varphi_{21}\,.
}
All other 1d operators belonging to the free sector are powers of $\cO^{(\frac12, \frac12)}_{\text{free}}(\varphi,y)$: 
\es{general1dfree}{
\cO^{(j,j)}_{\text{free}}(\varphi,y) = [\cO^{(\frac12, \frac12)}_{\text{free}}(\varphi,y)]^{2j}\,.
}
It follows that all free theory correlations functions can be computed using Wick contractions with the propagator \eqref{freeProp}. For the two and three point functions, we find
\es{free2pt}{
\langle\cO^{(j_1,j_1)}_{\text{free}}(\varphi_1, y_1)\cO^{(j_2,j_2)}_{\text{free}}(\varphi_2, y_2)\rangle= \delta_{j_1,j_2}(2j_1)!\left(\frac{3}{8 \pi r}\braket{y_1,y_2}\sgn(\varphi_{21} )\right)^{2j_1}
}
and, when $j_1$, $j_2$, $j_3$ obey the triangle inequality, 
\es{free3pt}{
\langle\cO^{(j_1,j_1)}_{\text{free}}(\varphi_1, y_1)\cO^{(j_2,j_2)}_{\text{free}}(\varphi_2, y_2)\cO^{(j_3,j_3)}_{\text{free}}(\varphi_3, y_3)\rangle = j_{123}! j_{231}! j_{321}! {2j_1 \choose j_{123}}{2j_2 \choose j_{231}}{2j_3 \choose j_{312}} \\ \times  \left(\frac{3}{8 \pi r} \sgn \: \varphi_{32} \braket{y_1,y_2}\right)^{j_{123}}\left(\frac{3}{8 \pi r} \sgn \: \varphi_{32} \braket{y_2,y_3}\right)^{j_{321}}\left(\frac{3}{8 \pi r}\sgn \: \varphi_{13} \braket{y_3,y_1}\right)^{j_{312}}\,.
}
Rescaling the $\cO^{(j,j)}_{\text{free}}$ by a positive factor in order to match \eqref{twistNorm} and comparing with \eqref{twistNorm3}, we extract the OPE coefficients 
\es{OPEfree}{
 \lambda_{(j_1, j_1), (j_2, j_2), (j_3,j_3)}^{\text{free}} =   \frac{ j_{123}! j_{231}! j_{321}!}{\sqrt{(2j_1)! (2j_2)! (2j_3)!}}  {2j_1 \choose j_{123}}{2j_2 \choose j_{231}}{2j_3 \choose j_{312}} \,.
}

\subsubsection{Interacting Sector}
\label{INTERACTING}

Let us now discuss operators in the interacting sector in increasing order of the number of ${\cal X}$'s they are built from.  The interacting sector cannot have any operators linear in ${\cal X}$, because such operators would have originated from $\Delta = 1/2$ operators in 3d, which are free.  So, the first non-trivial operator in the interacting sector must involve two ${\cal X}$'s. It must also be orthogonal to the free theory operator that is quadratic in ${\cal X}$, namely ${\cal O}^{(1,1)}_{\text{free}}$ defined in \eqref{general1dfree}. From this, one can show that such an operator is proportional to
\es{int2}{
{\cal O}^{(1,1)}_{\text{int}}(\varphi, y) = (\tr {\cal X}^2) (\varphi, y) - \frac{1}{3} (\tr {\cal X})^2 (\varphi, y) \,.
}

Next, we consider operators with three ${\cal X}$'s. It can be shown that the interacting sector contains only one such operator, which by assumption must be orthogonal to the operator ${\cal O}^{(\frac 32, \frac 32)}_{\text{free}}$ of the free sector as well as the operator ${\cal O}^{(\frac 12, \frac 12)}_{\text{free}} {\cal O}^{(1,1)}_{\text{int}}$ of the mixed sector. It follows that this operator in the interacting sector is proportional to
\es{int3}{
{\cal O}^{(\frac 32,\frac 32)}_{\text{int}}(\varphi, y) = (\tr {\cal X}^3) (\varphi, y) - \, \left( \tr  {\cal X}^2 \tr {\cal X} \right) (\varphi, y) + \frac29 (\tr {\cal X})^3(\varphi, y)   \,.
}

Next, we can construct operators with four ${\cal X}$'s.  It can be shown that the interacting sector contains two such operators. One of them has $j=2$ and is ${\cal O}^{2, 2}_\text{int} = ({\cal O}^{(1,1)}_{\text{int}})^2$.  The other has $j=0$, and is given by: 
\es{int20}{
{\cal O}^{(2,0)}_\text{int}(\varphi) = \epsilon^{ac} \epsilon^{bd} {\cO}^{(1,1)}_{\text{int}, ab}(\varphi) {\cO}^{(1,1)}_{\text{int}, cd}(\varphi)
- \frac{3 (2 \pi- 7)}{2 (\pi - 3) ( 4\pi r)^2}\,,
}
where here we have used explicit $SU(2)_F$ indices. The second term in the above expression ensures that this operator is orthogonal to the unit operator.  It is straightforward to continue and construct operators with more than four ${\cal X}$'s.

We can now use the propagator \eqref{Xprop} and the matrix model partition function \eqref{ABJMGen} to compute two and three point functions. For instance, for ${\cal O}^{(1,1)}_{\text{int}}(\varphi, y)$ we compute the two point function
\es{twisted-2pt-11ABJM}{
\braket{ \cO_\text{int}^{(1,1)}(\varphi_1, y_1)  \cO_\text{int}^{(1,1)}(\varphi_2, y_2)} 
&= \frac{\braket{y_1, y_2}^2}{Z_{\text{ABJM}_{3,1}} (4 \pi r)^2 } \int d^3 \sigma \: Z_{\text{ABJM}_{3,1}}^\sigma \left(1 + \sum_{\alpha<\beta} \sech^2(\pi \sigma_{\alpha \beta}) \right) \\
  &= \frac{10 \pi -31}{2 (\pi -3) (4 \pi r)^2}
\braket{y_1, y_2}^2 \,.
}
A similar calculation gives the three point function 
\es{twisted-3pt-11ABJM}{
\braket{ \cO_\text{int}^{(1,1)}(\varphi_1, y_1)  \cO_\text{int}^{(1,1)}(\varphi_2, y_2) \cO_\text{int}^{(1,1)}(\varphi_3, y_3)}  =
 \frac{10\pi -31}{(\pi - 3)(4 \pi r)^3} 
 \\ \times \text{sgn} \: \varphi_{21} \, \text{sgn} \: \varphi_{32} \,\text{sgn} \: \varphi_{13} \braket{y_1, y_2} \braket{y_2, y_3} \braket{y_3, y_1} \,.
}
Rescaling ${\cal O}_\text{int}^{(1, 1)}$ by a positive factor in order to match \eqref{twistNorm} and comparing with \eqref{twistNorm3}, we extract the OPE coefficient
\es{OPEcoeff}{
\lambda_{(1, 1), (1, 1), (1,1)} =  \sqrt{ \frac{8 (\pi -3)}{10 \pi -31} }\,.
}
Two other Higgs branch operators appear in the $\cO_\text{int}^{(1,1)}\times \cO_\text{int}^{(1,1)}$ OPE: $\cO_\text{int}^{(2,0)}$ and $\cO_\text{int}^{(2,2)}$. Performing the analogous calculation for these other operators yields the OPE coefficients 
\es{OPEcoeffs2}{
  \lambda _{(1, 1), (1, 1), (2,2)} &= \sqrt{\frac{2 (\pi -3) (840 \pi -2629)}{5 (10 \pi-31 )^2}} \,, \\
 \lambda _{(1, 1), (1, 1), (2,0)} &= \sqrt{\frac{3888+\pi  (420 \pi -2557)}{3 (10 \pi -31)^2}} \,.
}
As a consistency check, these OPE coefficients satisfy the relations
\es{2222cross}{
3\lambda^2_{(1, 1), (1, 1), (1,1)} - 5 \lambda^2_{(1, 1), (1, 1), (2, 2)} + 6\lambda^2_{(1, 1), (1, 1), (2,0)} + 6 = 0\,,
}
which were derived in \cite{Chester:2014mea}\footnote{The normalization of the OPE coefficients here differs from that in \cite{Chester:2014mea}. See Appendix~\ref{4point} for the relation between the two.} by applying crossing symmetry to the four point function of the 1d theory, which we review in Appendix \ref{4point}. We can convert these OPE coefficients to the conventions of \cite{Chester:2014fya} as described in Appendix \ref{4point}, to find 
\es{OPEFinal}{
 \bar\lambda_{2, 2, 1,1}^2 =&   \frac{16 (\pi -3)}{10 \pi -31}\,,\\
 \bar\lambda _{2,2,2,2}^2 =& \frac{16 (\pi -3) (840 \pi -2629)}{15 (10 \pi-31 )^2}\,, \\
 \bar\lambda _{2,2,2,0}^2 =& 16\frac{3888+\pi  (420 \pi -2557)}{3 (10 \pi -31)^2} \,.
}
We have used these coefficients in Figure~\ref{BBounds} to compare to an improved version of the conformal bootstrap bounds of \cite{Chester:2014fya}. 

We also computed the OPE coefficients for Higgs branch operators in the $\cO_\text{int}^{(1,1)}\times \cO_\text{int}^{(2,2)}$ and $\cO_\text{int}^{(2,2)}\times \cO_\text{int}^{(2,2)}$ OPEs. These expressions are more complicated, so we relegate them to Appendix \ref{More}.

\subsection{BLG$_3$}
\label{1dBLG}

As explained above, the 1d theory corresponding to the BLG theory requires a generalization of \cite{Dedushenko:2016jxl}.  If we are not interested in correlation functions of operators arising from monopole operators in 3d, we conjecture that we can simply insert \eqref{insertion} into \eqref{1dSector} and compute correlation functions of gauge-invariant operators built from $Q$ and $\tQ$.  For the BLG$_3$ theory, this conjecture produces the 1d theory
\es{BLGPartitionFunction1d}{
Z_\text{BLG$_3$} = \frac{1}{16\pi^2} \int d^2\sigma_\pm  e^{\frac{6 i  \sigma_+\sigma_-}{\pi}} \left( \frac{ \sinh(\sigma_++\sigma_-) \sinh( \sigma_+-\sigma_-)}{ \cosh(\sigma_+)\cosh(\sigma_-)}\right)^2 \int D\widetilde Q_{\alpha}^{\: \:\dot \beta}D{Q}_{\dot \beta}^{\: \: \alpha} e^{-S_\text{BLG$_3$}},
}
with
\es{BLGAction1d}{
S_\text{BLG$_3$} = -4r \int_{-\pi}^\pi d\varphi \:  \left[ \pi\widetilde{Q}_{\alpha}^{\: \: \dot \beta} \partial_\varphi Q_{\dot \beta}^{\: \: \alpha} + \sigma_-\widetilde{Q}_{1}^{\: \:\dot1} Q_{\dot1}^{\: \: 1} - \sigma_+\widetilde{Q}_{1}^{\: \: \dot2} Q_{\dot2}^{\: \: 1}  + \sigma_+\widetilde{Q}_{2}^{\: \: \dot1} Q_{\dot1}^{\: \: 2} - \sigma_-\widetilde{Q}_{2}^{\: \: \dot2} Q_{\dot2}^{\: \: 2} \right]\,,
}
where $\alpha,\beta$ and $\dot \alpha,\dot \beta$ are fundamental indices for each gauge group,  $ Q_{\dot \beta}^{\: \: \alpha}$ and $\widetilde Q_{\alpha}^{\: \: \dot \beta}$ correspond to the bifundamental hypermultiplets, and $\sigma_\pm $ are the same integration variables as in \eqref{partitionBLGVal}.  (Eq.~\eqref{partitionBLGVal} is obtained after integrating out $Q$ and $\tQ$ in \eqref{BLGPartitionFunction1d}.)   

We can rewrite the action in terms of the mass matrix-like quantity
\es{mass}{
M^{\: \:  \dot \beta}_{ \alpha} = 
\begin{pmatrix} 
\sigma_- \: & \: -\sigma_+ \\
\sigma_+ \: & \: - \sigma_-
\end{pmatrix}\,,
}
to read off the propagator
\es{BLGprop}{
\braket{Q_{\dot \beta}^{\: \: \alpha}(\varphi_1,y_1) \widetilde{Q}_{\gamma}^{\: \: \dot \delta}(\varphi_2,y_2)}_{\sigma} = -\delta_{\dot \beta}^{\: \: \dot \delta} \delta_{\gamma}^{\: \: \alpha}  \: \frac{\text{sgn} \: \varphi_{12} + \tanh(\pi M_\alpha^{\:\: \dot \beta})}{8 \pi r} e^{- M_\alpha^{\:\: \dot \beta} \varphi_{12}} \,,
}
where there is no sum over the gauge indices. We then compute correlation functions as 
\es{BLGgeneralCorrelation}{
\braket{\cO_1(\varphi_1,y_1) \cdots \cO_n(\varphi_n,y_n)} =& \frac{1}{Z _\text{BLG$_3$}} \int d^2\sigma_\pm  \: Z^\sigma _\text{BLG$_3$}
 \braket{\cO_1(\varphi_1,y_1) \cdots \cO_n(\varphi_n,y_n)}_\sigma \,,\\
Z^\sigma _\text{BLG$_3$}=& \frac{e^{\frac{6 i \sigma_+ \sigma_-}{\pi}}}{64 \pi^2}   \left(
\frac{\sinh(\sigma_+ + \sigma_-)\sinh(\sigma_+ - \sigma_-)}{\cosh^2(\sigma_+) \cosh^2(\sigma_-)}
 \right)^2 \,,
}
where $\braket{\cO_1(\varphi_1,y_1) \cdots \cO_n(\varphi_n,y_n)}_\sigma$ is the correlation function for the Gaussian theory \eqref{BLGAction1d} at fixed $\sigma$, given in \eqref{BLGprop}.

Since the 1d theory \eqref{BLGAction1d} arises from an ${\cal N} = 8$ SCFT, it must have a flavor $SU(2)_F$ symmetry.  Indeed, it can be checked that such a symmetry is present and that $(Q_{\dot \beta}{}^\alpha, \epsilon^{\alpha\gamma} \epsilon_{\dot \beta \dot \delta} \tQ_\gamma{}^{\dot \delta})$ form a doublet.  It is thus convenient to combine the $2\times2 $ matrices $Q$ and $\tQ$ into the matrix 
 \es{bfQDef}{
  \textbf{Q}(\varphi,  y) &=
\begin{pmatrix}
Q_{\dot1}^{\:\:1} y^1 - \widetilde{Q}_2^{\:\:\dot2} y^2 \: & \: Q_{\dot1}^{\:\:2}  y^1+ \widetilde{Q}_2^{\:\:\dot1} y^2 \\
Q_{\dot2}^{\:\:1}  y^1 + \widetilde{Q}_1^{\:\:\dot2} y^2 \: & \: Q_{\dot2}^{\:\:2}  y^1 - \widetilde{Q}_1^{\:\:\dot1} y^2
\end{pmatrix}\,,
 }
where $y^a$ are our usual $SU(2)_F$ polarization variables.

Let us see what gauge-invariant operators we can construct using an increasing number of ${\bf Q}$'s.  There are no gauge-invariant operators built from only one ${\bf Q}$.  With two ${\bf Q}$'s we can construct operators, which taken together have $SU(2)_F$ spin $j=1$ and can be written compactly as
\es{flavMat}{
\cO^{(1,1)}_{\text{BLG}_3}(\vphi, y) &= \det \textbf{Q}(\varphi,  y)\,.
}
With three ${\bf Q}$'s we again cannot construct any gauge-invariant operators.  With four ${\bf Q}$'s we can construct two operators: one with $j=2$, namely $\cO^{(2,2)}_{\text{BLG}_3} = (\cO^{(1,1)}_{\text{BLG}_3} )^2$, and one with $j=0$, namely
\es{int20BLG}{
\cO^{(2,0)}_{\text{BLG}_3} (\varphi)= \epsilon^{ac} \epsilon^{bd} \cO^{(1,1)}_{\text{BLG}_3, ab}(\varphi) \cO^{(1,1)}_{\text{BLG}_3, cd}(\varphi)
- \frac{3 (2 \pi- 7)}{2 (\pi - 3) (4 \pi r)^2}\,,
}
where here we have again used explicit $SU(2)_F$ indices and have included a second term to ensure that it is orthogonal to the unit operator.  It is straightforward to proceed further using five ${\bf Q}$'s and higher.

The operators constructed so far, namely $\cO^{(1,1)}_{\text{BLG}_3}$, $\cO^{(2, 2)}_{\text{BLG}_3}$, and $\cO^{(2, 0)}_{\text{BLG}_3}$, match a subset of the operators we constructed in Section~\ref{INTERACTING} for the interacting sector of the ABJM$_{3, 1}$ theory, namely  $\cO_\text{int}^{(1,1)}$, $\cO_\text{int}^{(\frac 32,\frac 32)}$, $\cO_\text{int}^{(2,2)}$, $\cO_\text{int}^{(2,0)}$.  We were not able to construct the BLG$_3$ analog of $\cO_\text{int}^{(\frac 32,\frac 32)}$ using only the ${\bf Q}$'s because this operator requires monopole operators.

Nevertheless, we can use the propagator \eqref{BLGprop} and the matrix model partition function \eqref{BLGgeneralCorrelation} to compute two and three point functions of the operators we were able to construct in the 1d theory \eqref{BLGAction1d}, and compare them to the analogous expressions from the interacting sector of the ABJM$_{3, 1}$ theory.  For instance, for ${\cal O}^{(1,1)}_{\text{BLG$_3$}}(\varphi, y)$ we compute the two point function
\es{twisted-2pt-11BLG}{
\braket{ \cO_{\text{BLG}_3}^{(1,1)}(\varphi_1, y_1)  \cO_{\text{BLG}_3}^{(1,1)}(\varphi_2, y_2)}&=
\frac{\braket{y_1, y_2}^2}{4Z _\text{BLG$_3$} (4 \pi r)^2 } \int d^2\sigma_\pm \: Z _\text{BLG$_3$}^\sigma 
(\sech^2(\sigma_-) + \sech^2( \sigma_+))\\
&= \frac{10 \pi -31}{8 (\pi -3) (4 \pi r)^2}
\braket{y_1, y_2}^2 \,.
}
A similar calculation gives the three point function 
\es{twisted-3pt-11BLG}{
\braket{ \cO_{\text{BLG}_3}^{(1,1)}(\varphi_1, y_1)  \cO_{\text{BLG}_3}^{(1,1)}(\varphi_2, y_2)  \cO_{\text{BLG}_3}^{(1,1)}(\varphi_3, y_3)} =
\frac{10\pi -31}{4(\pi - 3)(4 \pi r)^3} 
 \\ \times \text{sgn} \: \varphi_{21} \, \text{sgn} \: \varphi_{32} \,\text{sgn} \: \varphi_{13} \braket{y_1, y_2} \braket{y_2, y_3} \braket{y_3, y_1}  \,.
}
By comparing to \eqref{twistNorm} and \eqref{twistNorm3} (we rescale $\cO_{\text{BLG}_3}^{(1,1)}$ by a positive factor in order to match \eqref{twistNorm}), we extract the OPE coefficient
\es{OPEcoeffBLG}{
\lambda_{(1, 1), (1, 1), (1,1)} =   \sqrt{\frac{8 (\pi -3)}{10 \pi -31}}\,,
}
which agrees with \eqref{OPEcoeff} for the interacting sector of the ABJM$_{3, 1}$ theory. We can similarly check that the OPE coefficients of all the other Higgs branch operators that appear in the $\cO_\text{BLG$_3$}^{(1,1)}\times \cO_\text{BLG$_3$}^{(2,2)}$, $\cO_\text{BLG$_3$}^{(1,1)}\times \cO_\text{BLG$_3$}^{(2,2)}$, and $\cO_\text{BLG$_3$}^{(1,1)}\times \cO_\text{BLG$_3$}^{(2,2)}$ OPEs match those of ABJM$_{3, 1}$ theory, given in \eqref{OPEcoeffs2} and \eqref{OPEcoefficentsAll}.

\section*{Acknowledgments}

This work was supported in part by the US NSF under grant No.~PHY-1418069 and by the Simons Foundation grant No.~488651.

\appendix

\section{Four Point Function}
\label{4point}

When $\varphi_1 < \varphi_2 < \varphi_3 < \varphi_4$ and $j_1 \geq j_2$, the four point function of $(B,+)^{[00(2j)0]}$ operators $\cO^{(j,j)}(\varphi,y)$ can be decomposed in the s-channel as 
\es{twisted-4pt}{
&\braket{{\cal O}^{(j_1,j_1)}(\varphi_1,y_1) {\cal O}^{(j_2,j_2)}(\varphi_2,y_2) {\cal O}^{(j_3,j_3)}(\varphi_3, y_3) 
{\cal O}^{(j_4,j_4)}(\varphi_4, y_4)} = \braket{y_1,y_2}^{j_1+j_2} \braket{y_3, y_4}^{j_3+j_4} \\ 
&\qquad\qquad\qquad\times\left[\frac{\braket{y_1, y_4}}{\braket{y_2, y_4}}\right]^{j_{12}}
\left[\frac{\braket{y_1,y_3}}{\braket{y_1, y_4}}\right]^{j_{34}} 
\sum_{\Delta=j_1-j_2}^{j_1+j_2}\sum_{j=j_1-j_2}^\Delta \frac{t_j(w)}{4^\Delta} {\bar \lambda}_{2j_1, 2j_2,\Delta,j} {\bar \lambda}_{2j_3, 2j_4,\Delta,j} \,,
}
where $w$ is the $SU(2)_F$ cross-ratio
\es{su2-cross-ratio}{
w = \frac{\braket{y_1,y_2} \braket{y_3,y_4}}{\braket{y_1,y_3}\braket{y_2,y_4}}\,,
}
and we have normalized the OPE coefficients ${\bar \lambda}$ as in \cite{Chester:2014fya}. For $j_1 = j_2$ we have the extra constraint $\Delta+j \in \text{Even}$, because scalar Higgs branch operators can only appear in the symmetric product of identical operators.
The function $t_j(w)$ obeys the eigenvalue equation:
 \es{tDiff}{
   (1-w) w^2 \frac{d^2 t_j}{dw^2}  +(j_{34} - j_{12} - 1) w^2 \frac{d t_j}{dw} + j_{12} j_{34} w t_j  = j(j+1) t_j \,.
 }
Up to normalization, the regular solution can be written in terms of the Jacobi polynomials $P_n^{(a, b)}(x)$ as 
 \es{GotT}{
  t_j (w) = w^{j_{34}} P_{j + j_{34}}^{(j_{12} - j_{34} , -j_{12}-j_{34})} \left(\frac {2}{w} -1 \right) \,.
 }
Note that when this expression is plugged into \eqref{twisted-4pt}, the total expression is a polynomial in the $y$'s. The OPE coefficients $\bar\lambda$ in \eqref{twisted-4pt} are related to $\lambda$ in \eqref{twistNorm3} by 
\es{OPEconversion}{
\lambda_{(j_1, j_1),(j_2, j_2),(\Delta,j)} \lambda_{(j_3, j_3),(j_4, j_4),(\Delta,j)} =  \lim_{w \to 0} \frac{w^{j} t_j(w)}{4^\Delta} {\bar \lambda}_{2j_1,2j_2,\Delta,j}{\bar \lambda}_{2j_3,2j_4,\Delta,j} \,,
}
where here we do not sum over repeated indices.

We can also write the four point function in the t-channel by exchanging $(1\leftrightarrow3)$ in \eqref{twisted-4pt}. Equating the s- and t-channels yields the following finite set of crossing equations
\es{crossConstraints}{
&\braket{1111}:
4{\bar \lambda}^2_{2,2,1,1} - 5 {\bar \lambda}^2_{2,2,2,2} +{\bar \lambda}^2_{2,2,2,0} + 16 = 0 \,. \\
&\braket{2222}:
\begin{cases}
64 {\bar \lambda} _{4,4,1,1}^2+48{\bar \lambda} _{4,4,2,2}^2+4 {\bar \lambda}
   _{4,4,3,1}^2-16{\bar \lambda} _{4,4,3,3}^2+ 3{\bar \lambda} _{4,4,4,2}^2-60   {\bar \lambda} _{4,4,4,4}^2 = 0\,, \\
   32 {\bar \lambda} _{4,4,2,2}^2+2{\bar \lambda} _{4,4,4,2}^2+9 {\bar \lambda}
   _{4,4,4,4}^2-16 {\bar \lambda} _{4,4,2,0}^2- 20
   {\bar \lambda} _{4,4,3,3}^2-{\bar \lambda} _{4,4,4,0}^2-256 = 0 \,.  \end{cases} \\
&\braket{1212}:
16 {\bar \lambda} _{2,4,1,1}^2+4{\bar \lambda} _{2,4,2,1}^2+4 {\bar \lambda} _{2,4,2,2}^2+{\bar \lambda}
   _{2,4,3,1}^2+{\bar \lambda} _{2,4,3,2}^2-14 {\bar \lambda} _{2,4,3,3}^2 = 0 \,.\\
&\braket{2112}:
\begin{cases}
  5 {\bar \lambda} _{2,4,3,1}^2+80 {\bar \lambda} _{2,4,1,1}^2+20{\bar \lambda} _{2,4,2,1}^2-12 {\bar \lambda} _{2,2,2,0} {\bar \lambda}
   _{4,4,2,0}+72 {\bar \lambda} _{2,2,1,1} {\bar \lambda}
   _{4,4,1,1}\\
   -42{\bar \lambda} _{2,2,2,2} {\bar \lambda}
   _{4,4,2,2}-192=0 \,, \\
     3 {\bar \lambda} _{2,4,3,2}^2+12 {\bar \lambda} _{2,4,2,2}^2-14 {\bar \lambda} _{2,2,2,2}
   {\bar \lambda} _{4,4,2,2}\\
   -8 {\bar \lambda} _{2,2,1,1}
   {\bar \lambda} _{4,4,1,1}+4 {\bar \lambda} _{2,2,2,0}
   {\bar \lambda} _{4,4,2,0} +64= 0 \,, \\
   15 {\bar \lambda} _{2,4,3,3}^2 -4 {\bar \lambda} _{2,2,2,2}
   {\bar \lambda} _{4,4,2,2}-16 {\bar \lambda} _{2,2,1,1}
  {\bar \lambda} _{4,4,1,1}-4 {\bar \lambda} _{2,2,2,0}
   {\bar \lambda} _{4,4,2,0} -64= 0 \,.
   \end{cases}
}

\section{OPE Coefficients in $\cO^{(1,1)}\times \cO^{(2,2)}$ and $\cO^{(2,2)}\times \cO^{(2,2)}$}
 \label{More}
 In Sections~\ref{1dABJM} and~\ref{1dBLG} we described how to compute OPE coefficients of Higgs branch operators that lie in the 1d topological sector of the ABJM$_{3, 1}$ and BLG$_3$ theories, respectively. In \eqref{OPEcoeff} and \eqref{OPEcoeffs2} we list the OPE coefficients of Higgs branch operators that appear in the $\cO^{(1,1)}\times \cO^{(1,1)}$ OPE\@. By a similar calculation, we can compute the OPE coefficients of Higgs branch operators that appear in the $\cO^{(1,1)}\times \cO^{(2,2)}$ and $\cO^{(2,2)}\times \cO^{(2,2)}$ OPEs for either theory. We find
 \es{OPEcoefficentsAll}{
 \lambda _{(2, 2), (2, 2),(1,1)} &= \sqrt{\frac{32 (\pi -3)}{10 \pi -31}}\,, \\
 \lambda _{(2, 2), (2, 2),(2,2)} &= \sqrt{\frac{40 (521767-166320 \pi )^2 (\pi -3)}{49 (840 \pi
   -2629)^3}}\,, \\
 \lambda _{(2, 2), (2, 2),(3,3)} &= \sqrt{\frac{14400 (\pi -3) (9520 \pi -29877)}{7 (2629-840 \pi )^2}}\,, \\
 \lambda _{(2, 2), (2, 2),(4,4)} &= \sqrt{\frac{30 (\pi -3) (4583040 \pi -14394049)}{7 (2629-840 \pi
   )^2}}\,, \\
 \lambda _{(2, 2), (2, 2),(4,2)} &= \sqrt{\frac{960 (\pi -3) (4447712646+35 \pi  (12972960 \pi
   -81205777))}{49 (840 \pi -2629)^3}}\,, \\
 \lambda _{(2, 2), (2, 2),(4,0)} &= \sqrt{\frac{4 (\pi -3) (\pi  (8530357644+35 \pi  (8648640 \pi
   -79544233))-8707129344)}{(2629-840 \pi )^2 (3888+\pi  (420 \pi -2557))}}\,, \\
 \lambda _{(2, 2), (2, 2),(3,1)} &= \sqrt{\frac{64 (\pi -3) (2675592+5 \pi  (55440 \pi
   -344503))}{(2629-840 \pi )^2 (10 \pi -31)}}\,, \\
 \lambda _{(2, 2), (2, 2),(2,0)} &= \sqrt{\frac{(847584+\pi  (90720 \pi -554797))^2}{3 (2629-840 \pi
   )^2 (3888+\pi  (420 \pi -2557))}}\,, \\
 \lambda _{(1, 1), (2, 2),(1,1)} &= \sqrt{\frac{2 (\pi -3) (840 \pi -2629)}{5 (31-10 \pi )^2}}\,, \\
 \lambda _{(1, 1), (2, 2),(2,2)} &= \sqrt{\frac{32 (\pi -3)}{10 \pi -31}}\,, \\
 \lambda _{(1, 1), (2, 2), (3,3)} &= \sqrt{\frac{45 (\pi -3) (9520 \pi -29877)}{7 (10 \pi -31) (840 \pi
   -2629)}}\,, \\
 \lambda _{(1, 1), (2, 2),(3,1)} &= \sqrt{\frac{4(\pi -3) (2675592+5 \pi  (55440 \pi -344503))}{5
   (31-10 \pi )^2 (840 \pi -2629)}} \,.
}
As a consistency check, these OPE coefficients satisfy the crossing relations \eqref{crossConstraints} after we convert from $\lambda$ to $\bar\lambda$ using \eqref{OPEconversion}.
 
\bibliographystyle{ssg}
\bibliography{Duality}

\end{document}